# INGOT Wavefront Sensor: Simulation of Pupil Images

Valentina Viotto[ab], Elisa Portaluri[ab], Carmelo Arcidiacono[ab], Maria Bergomi[ab], Simone Di Filippo[bc], Davide Greggio[ab], Kalyan Radhakrishnan[ab], Marco Dima[ab], Jacopo Farinato[ab], Demetrio Magrin[ab], Luca Marafatto[ab] and Roberto Ragazzoni[ab]

[a]INAF – Osservatorio Astronomico di Padova, Vicolo dell'Osservatorio 5, 35122, Padova, Italy
[b]ADONI – ADaptive Optics National laboratory in Italy
[c]Università degli Studi di Padova – Dipartimento di Fisica e Astronomia, Vicolo dell'Osservatorio 3, 35122, Padova, Italy

## ABSTRACT

The ingot wavefront sensor (I-WFS) has been proposed, for ELT-like apertures, as a possible pupil plane WFS, to cope with the geometrical characteristics of a laser guide star (LGS).

Within the study and development of such a WFS, on-going in the framework of the MAORY project, the final purpose of the I-WFS simulation is to estimate its performance in terms of wavefront aberration measurement capability. The first step of this analysis is to translate incoming wavefronts into the three pupil images, produced by the optical system. The intrinsic geometrical characteristics of the ingot optical element, designed to be coupled with the LGS elongated image, make the system conceptually different with respect to other pupil WFSs (like the Pyramid WFS, P-WFS) also in terms of the simulation technique to be selected, within the ones which can be found in literature.

In this paper, we aim to report the considerations and derivations which led to the selection of a ray-tracing method for ingot pupil images simulation, and the geometrical assumptions and approach made to optimize the computing time.

**Keywords:** Laser guide star, wavefront sensing

## 1. INTRODUCTION

The next generation extremely large telescopes MCAO modules are going to be fed by Laser Guide Stars[1]. These artificial sources, have been developed in the past decades to complement Natural Guide Stars (NGSs), aiming at a large sky coverage, also in star poor regions. Despite their diffusion already in the 8 to 10 m class telescopes, these sources show major differences with respect to the natural stars, which we can approximate as point-like sources. In fact, the "smallest" LGSs are produced exciting, with a laser beacon, Sodium atoms populating a peculiar atmospheric layer at 90 km altitude, which, having a non-negligible thickness, produces a cigar-shaped source. This geometric peculiarity of the LGS, which, on top of this, is seen under different angles by each of the sub-apertures in a giant telescope, produces artificial effects on the signal retrieved by a classical wavefront sensor (WFS). These effects can be, at least, partially, calibrated, like the cone effect and the spot truncation, but they leave a signature in the total noise budget of the AO system.

To minimize this residual noise, which gets more and more impacting while the size of the telescope entrance pupil increases, we proposed a pupil plane WFS, designed on the actual shape of the LGS source: the Ingot wavefront sensor[2]. With the goal of evaluating the expected performance of this innovative sensor, compared with a classical Shack-Hartmann wavefront sensor (SH-WFS[3]), we started the development of a simulation tool (whose development is reported in [4]), initially following an approach analogous to what was done in the past for other WFSs. Like the peculiar shape of the LGS source pushed us to develop a dedicated wavefront sensor, the I-WFS, in turn, seems to require a different simulation approach, which takes into account its geometrical characteristics. Here we discuss the trade-off, carried out during the I-WFS simulator conceptual definition and development to find the best compromise between the possible approaches, including Fourier transform and ray tracing, to describe, at the best of our possibilities, the behavior of this new WFS.

## 2. WHY THE INGOT?

The reason to study a new kind of wavefront sensor lays, on one side, on the peculiarity of the Sodium LGS. This kind of source, unlike a real star, is intrinsically extended, elongated and with a given orientation in the 3D space and its actual shape convolves with the effect of diffraction and turbulence, to produce, in the focal plane area, a 3D image.

In Viotto et al. [5], we discussed the main reasons for which the development of a wavefront sensor, designed and optimized for the LGS source case, was considered a mandatory approach to overcome the limitations that some "classical" WFSs, like the Shack-Hartmann WFS and the Pyramid WFS (P-WFS[6]), originally devised for point-like sources, present.

We recall them here, as a reference, in the following list:

- the LGS is located at a finite distance from the observer;
- the LGS actual image forms on a plane, which is tilted with respect to the optical axis (in focal plane WFSs the spot is only partially in focus + truncation);
- the sodium layer thickness evolves[8], and so does the LGS elongation;
- we want to select a pupil plane WFS approach to optimize the pixels occupation;
- we want to consider by design differential LGS shape as seen from different sub-apertures.

## 3. INGOT WFS CONCEPTUAL DESIGN

To cope with the elongated image of the LGS, keeping a pupil-plane WFS approach, the basic idea was to have the optical element, discriminating the parts of the wavefront affected by different local first-derivative, positioned with the same orientation of the LGS in the image space, so to have a focal plane splitting component fitting the LGS 3D image both along the Z axis and in the XY plane, then acting as a 4-quad (or 6, or 3…) WFS. The first Ingot design included a six-faces (partially convex, partially concave, partially reflective, partially refractive) prism, so to split the LGS image into six areas, in the actual plane on which the image forms, sampling it in three parts along the star elongation direction in the XY plane and in 2 parts (like a four-quadrant) in the other direction (see Figure 1).

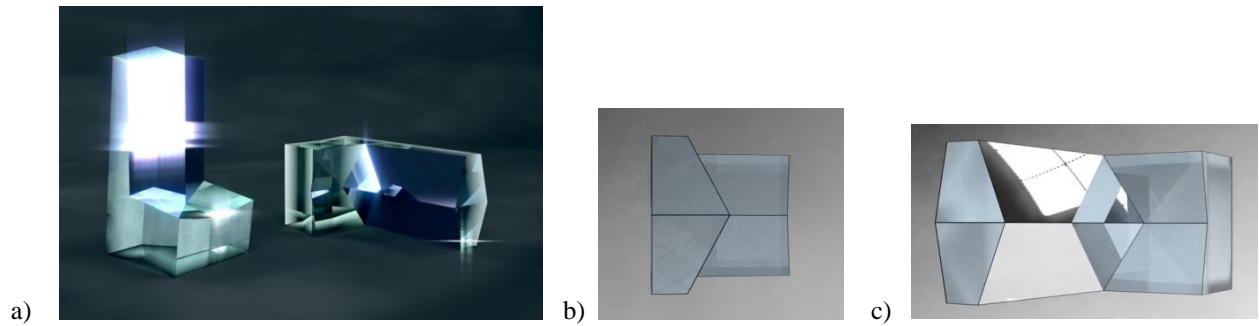

Figure 1. 6-faces Ingot prism design (a), and its projection, as seen from different sub-apertures on the entrance pupil (b, c). One can see that in case b), the sub-aperture close to the LGS launcher will only illuminates four out of the six faces of the prism, with a close-to-circular spot, like in a P-WFS, while in case c), the sub-aperture that receives the LGS light from the largest solid angle will then produce an elongated spot covering the six faces.

In this 6-faces prism Ingot, the signal was computed, from the re-imaged pupils, in a 6-quad-like style, using the light on the pupils produced by the four lateral faces to compute the wavefront derivative along both the axis, and the light in the central two faces only to compute the derivative signal in the direction orthogonal to the spot elongation (Y), under the assumption that the central faces mostly carry noise for the X signal[9].

The main limitation of the 6-faces design was identified in the rigidity of the prism size/shape, in contrast with the evolving elongation of the LGS image, due to both the pointing change in elevation and the intrinsic variability of the Sodium layer thickness. In fact, since the 6-faces Ingot is designed on the LGS image geometric characteristics, and they also include size, when the Sodium layer gets thicker, or the pointing elevation changes, the LGS doesn't match the Ingot shape anymore. In particular, some pupils could even get totally dark. This limitation was planned to be overcome equipping the WFS with a number of interchangeable Ingot prisms, to properly match the source projected elongation[5].

To reduce the complexity and make the Ingot WFS more adaptable to pointing and atmospheric conditions, an additional prism design was developed. In the 3-faces Ingot approach, in fact, depicted in Figure 2, the vertex of the prism is constrained to the lower edge of the Sodium layer, using the tip-tilt term, which is typically not used for LGS wavefront sensing (because of the LGS tilt indetermination problem[7]). In this way, most of the light illuminates the two reflecting faces, which split the first two beams, while the extreme side of the LGS image, produced by the lower edge of the Sodium layer, doesn't hit the prism at all and produces a third pupil, after the pupil re-imager (see Figure 3).

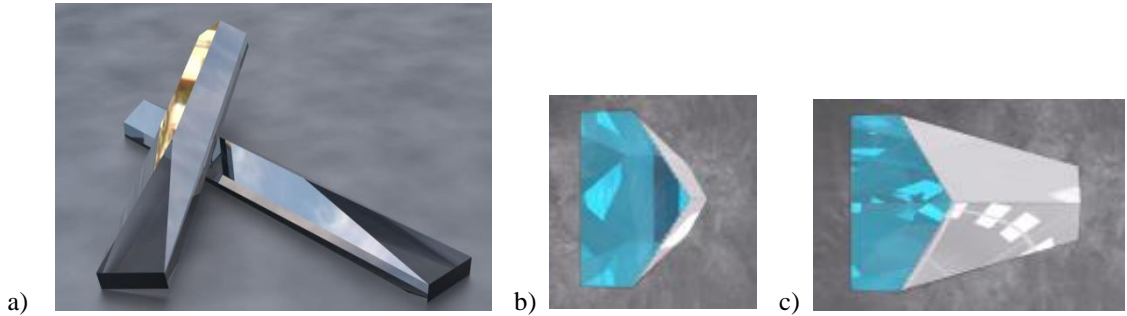

a) b) c)

Figure 2. 3-faces Ingot prism design (a), and its projection, as seen from different sub-apertures on the entrance pupil (b, c). One can see that in case b), the sub-aperture close to the LGS launcher will produce a close-to-circular spot, like in a P-WFS, while in case c), the sub-aperture that receives the LGS light from the largest solid angle will then produce an elongated spot. Both in cases b) and c) all the three pupils are illuminated.

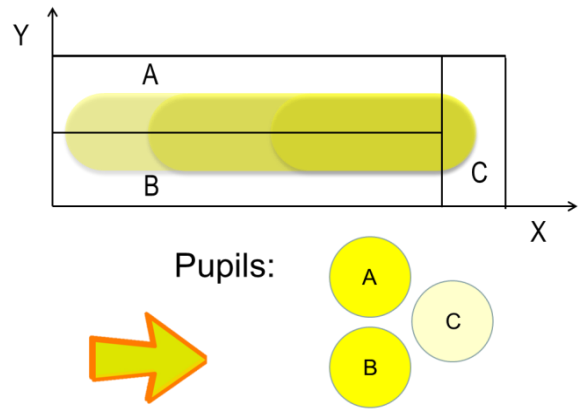

Figure 3. 3-faces Ingot prism scheme

The signals are, then computed as follows:

$$S_X = \frac{C}{<A+B>_{\sim seconds}} \cdot \gamma \quad \text{and} \quad S_Y = \frac{A-B}{<A+B>_{\sim seconds}} \quad \text{, where } \gamma = \frac{<A+B>_{\sim hour}}{<C>_{\sim hour}}$$

These figures include two normalization factors, obtained averaging the pupils counts over different time regimes. The first one ($<A+B>_{\sim seconds}$) is meant to deal with scintillation effect, while the second ($\gamma$) considers the "static" differential illumination of the pupil sub-apertures, due to the peculiar shape of the prism itself. However, the discussion on the best approach to compute the signals is beyond the scope of this proceeding.

We define the XY plane in Figure 3, on which the signals are computed, as the *Ingot prism main plane*. The Ingot prism main plane:

- includes the prism edge, which separates the two reflecting faces;
- forms equal angles with the two reflecting faces of the Ingot;
- on this plane, the centered vertical section of the LGS source is reimaged (and focused).

# 4. INGOT WFS SIMULATION APPROACHES DISCUSSION

To estimate the preliminary I-WFS performance, we investigated the possible simulation approaches, considering the peculiarities of the I-WFS with respect to more conventional WFSs, like the Shack-Hartmann and the Pyramid ones. As already discussed, the main differences lay in the 3D shape of the source and, consequently, of its image, produced by the telescope.

## 4.1 LGS source sampling

The first step of the simulations is to reproduce the source itself, so to be able to convert it into its image in the focal plane area (volume, in this case). The I-WFS was conceived in particular for the ELTs, where the telescope aperture size is getting less and less negligible, with respect to the distance and extension of the LGS source, and the spot truncation problem can have a larger impact on the AO error budget. In this configuration, the extended LGS is resolved by the telescope, and local spatial variations of the Sodium layer density cannot be neglected, since they may have an impact on the performance.

We selected geometric ray-tracing-like approach, in which the effect of diffraction is not included (except for the re-imaged size of the source, produced by the LGS sampling), since the ingot prism extends far from the nominal focal plane of the upstream optics (this would include fractional Fourier Transforms too).

We investigated three different representations of the LGS cigar:

- 3D sampling: the source is sampled with a 3D grid, in which differential intensities inside the LGS can be implemented both vertically (to match Na layer density spatial variations) and radially (a Gaussian profile can mimic Na atoms excitement radial profile). In this case you get $P^{III} = N_R^2 \pi \cdot N_V$ point-like sources, where $N_R$ and $N_V$ represent the number of dots along the radius of the LGS and the its elongation (optimal sampling will be different in the two directions), respectively.

- 2D sampling: the sampling only occurs on the plane, inside the LGS, which focuses on the Ingot prism main plane. Again, differential intensities can be applied. In this case you get $P^{II} = 2N_R \cdot N_V$ point-like sources.

- 1D sampling: the sampling occurs along the LGS axis only, which focuses on the Ingot prism roof edge. Then, each of the ray-traced points is convolved with the image of a Gaussian disk (a "slice of LGS") on the Ingot prism main plane, obtained through the telescope system. In this case you get $P^I = N_V$ point-like sources to be convolved, as a first-order approximation, with the same disk image.

Figure 4 graphically shows the different sampling approaches described above.

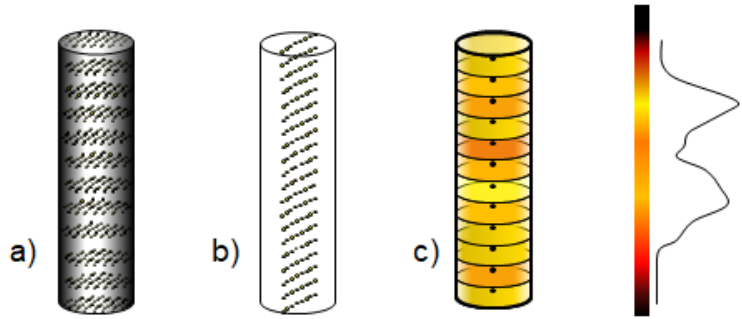

Figure 4. LGS sampling approaches: a) full source 3D sampling, differential radial intensity can be implemented, b) 2D grid, sampling the points focused on the Ingot main plane, c) sampling along the LGS axis, which is focused along the Ingot roof edge. Representative disk image needed. In any case, intensity shall follow representative Na-layer profile.

## 4.2 Point-like sources re-imaging onto the Ingot main plane

Once the source sampling approach is selected, we need to identify the best compromise to simulate the way each of the point-like sources, sampling the LGS, is re-imaged onto the Ingot main plane, where the discrimination between different tilt occurs.

**Fast Fourier Transform approach**

The classical Fourier approach, successfully implemented e.g. in P-WFS simulators, would be the first choice, being well known, reliable (under some assumptions), and requiring a modest computing time, when FFTs are used. Unfortunately, despite showing some similarities with the P-WFS, the I-WFS geometrical peculiarities have a non-negligible impact on the assumptions, usually made to use the Fourier approach.

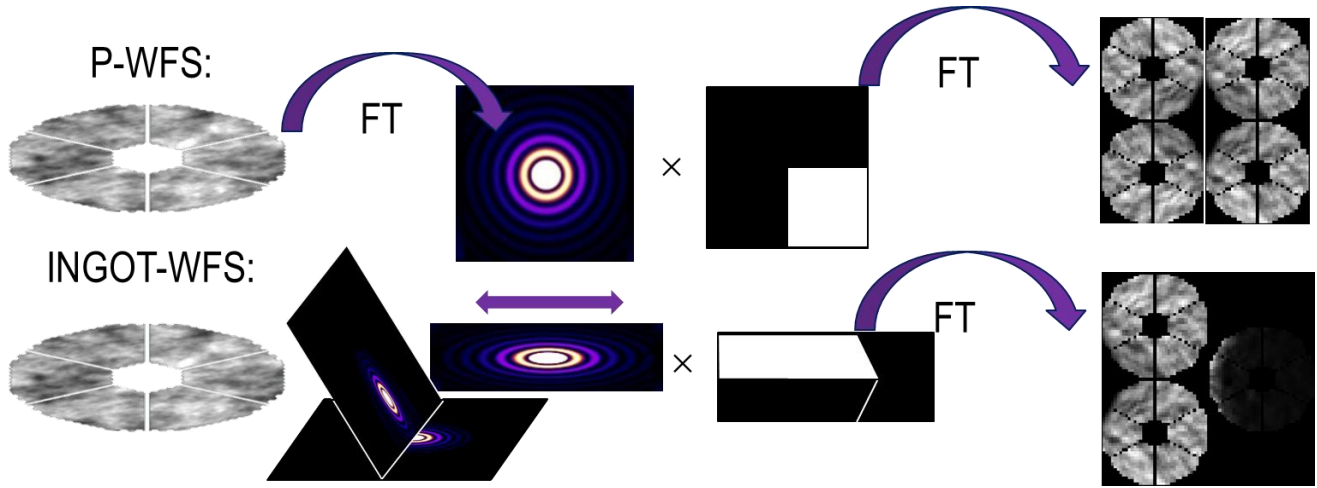

Figure 5. Fourier Transform approach: P-WFS case (top) is compared with I-WFS case (bottom).

The scheme in Figure 5, show, on the top, the pupil plane – focal plane (electric field) – pupil plane transformations, which can be obtained with the FFTs, in the P-WFS regime. To simulate the Pyramid prism, a phase mask (here represented, instead, with an amplitude mask, selecting one face of the pyramid) is multiplying the focal plane image, before it is anti-transformed in the pupil plane space. In the bottom part of the scheme, the analogous path is reported for the I-WFS. We highlight here the orientation of the Ingot main plane, on which the phase mask shall be applied, where the PSF is expected to be stretched along one axis.

Unfortunately, this approach, even including a PSF stretching to simulate the orientation of the Ingot main plane, cannot be representative of the real physics, for a couple of main reasons:

1. the orientation of the Ingot main plane, in fact, is close to be orthogonal to the focal plane on which objects, placed at infinity, are re-imaged (e.g. ~83deg for the E-ELT). In this direction, the PSF does not show the same (or stretched) pattern (see Figure 6 for a qualitative representation).

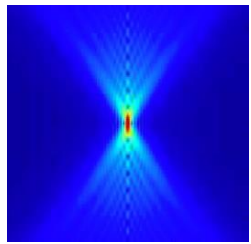

Figure 6. PSF pattern in a plane orthogonal to the focal plane

2. the vertex angle of the Pyramid prism, in the P-WFS, is very small (~1 deg). This is not the case for the Ingot prism, in which the reflective faces deviate from the Ingot main plane of ~30 deg. For this reason, we cannot approximate the Ingot with a common phase mask, since each of the sub-aperture will see the prism under a different orientation and would need a dedicated phase mask, projected onto the Ingot main plane. Figure 7 show, as an example, two projections of the Ingot faces onto the Ingot main plane, as seen by different sub-apertures in the entrance pupil.

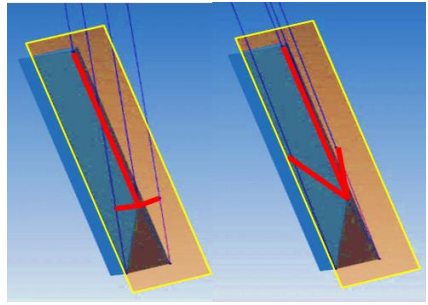

Figure 7. Projections of the Ingot faces onto the Ingot main plane (yellow rectangular region), as seen by different sub-apertures in the entrance pupil.

**Pure ray tracing approach**

For these reasons, we considered also a pure ray-tracing approach (already presented in Viotto et al. [5], for the 6-faces Ingot prism), in which each point-like source produces $N$ rays, where $N$ is the number of sub-apertures on the entrance pupil. The simulation considers as inputs the wavefront shape (which is actually retrieved from a turbulence profile, assuming frozen atmosphere and considering the cone effect), and the ELT pupil shape, sampled with a given density. Position of the laser launcher, LGS orientation with respect to the entrance pupil and Ingot prism sizes and angles are input parameters too. It is worth mentioning that, despite the fact that the pure ray tracing approach does not consider diffraction effects, their impact on the re-imaged LGS is expected to play a minor role, since the diffraction figure produced by a full ELT aperture is negligible in size, when compared to the actual diameter, in the sky, of the typical LGS cigar. This is not true in the case of a WFS sampling the pupil before the tilt optical sensing occurs (e.g. SH-WFS).

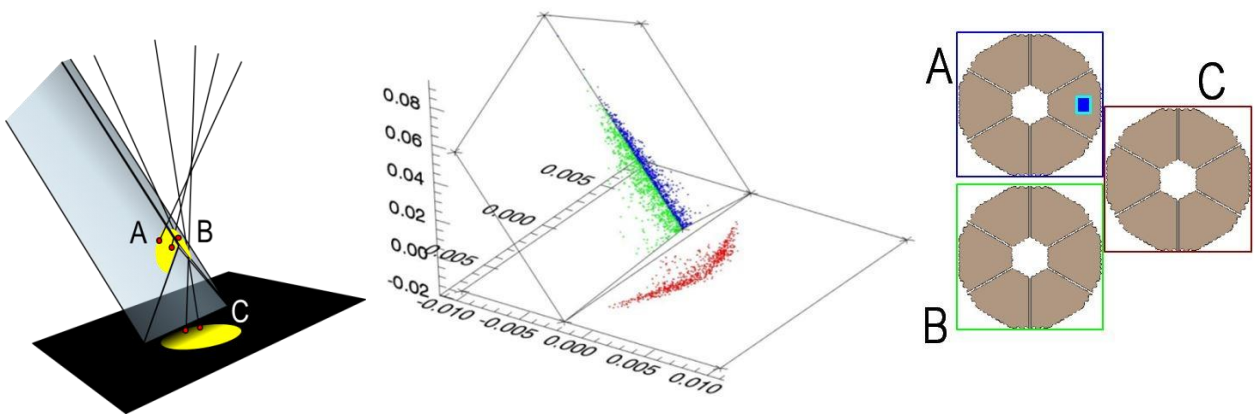

Figure 8. Ingot prism illuminated by incoming rays. *Left*: tracing scheme, *Center*: simulated intersections of rays with the Ingot faces, *Right*: each of the rays will add flux to only one of the three pupil images. A low cigar sampling and number of rays are reported for clarity sake. Units in the picture are meters, estimated for an ELT LGS projection, assuming a F/5 focal plane.

Figure 8 shows the scheme and the simulation of a bundle of rays, sampling a LGS and illuminating the prism, given an incoming Von Karman-like aberration. A limited number of rays and point-like sources is reported, for sake of clarity. The Ingot prism, simulated as composed by three quadrilaterals, defined in the 3D space, in the area of the telescope focal plane. Each pupil sub-aperture generates a ray for each point-like source illuminating it, perpendicular to the incoming wavefront, which intersects only one of the three faces of the prism. This allows to select which pupil the ray will illuminate.

The limitations of this approach consist in the low dynamic range of the pupil images, since each of the *NxP* (*N* being the number of sub-apertures and *P* the number of point-like source sampling the LGS) rays will only contribute to the flux of one of the three pupil images. For this reason, a dense sampling of the source may be needed, which, on the other hand, would require a long computing time.

**Hybrid ray tracing approach**

To try and overcome the limitations of the ray tracing approach we devised a 2-steps simulation scheme, that we can summarize as follows, starting from the LGS source 1D sampling, presented in Section 4.1:

1. System geometry calibration: to be performed without atmospheric aberrations, allows to model the system geometry, in terms of projections of the re-imaged point-like source and of the Ingot prism on the Ingot main plane.

    a. A LGS-slice, represented by a disk at 90 km altitude, with a thickness corresponding to the vertical sampling pitch and a radial intensity distribution representing the Sodium layer excitement profile, is sampled with a high density 3D grid, and each of the points in the grid is then re-imaged on the Ingot main plane. Since each sub-aperture sees the disk under a different angle, and this discrepancy is not-negligible, *N* images of the disk are computed. Such images are then parametrized so to get rid of the residual grid and minimize storage space needed.

    b. Parameters describing *N* Ingot projections on the Ingot main plane, as seen from each of the sub-apertures, are stored too.

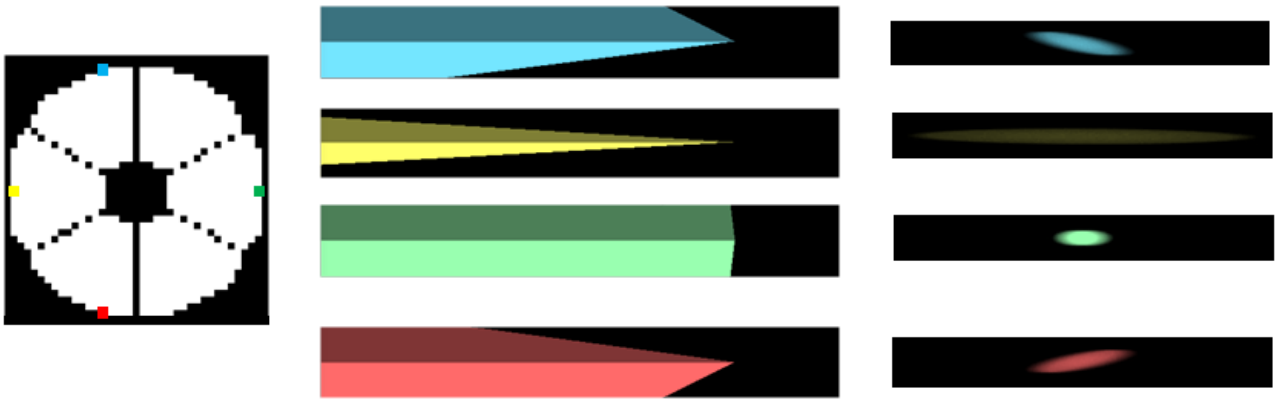

Figure 9. Ingot projections (center) and disk images (right) on the Ingot main plane, as seen from different sub-aperture, in the E-ELT geometry case. The laser launcher is located on the left of the yellow sub-aperture.

   As reported in Figure 9, the Ingot projections and disk images geometry, on the Ingot main plane, can vary very much one with respect to the other, depending on the sub-aperture.

2. Now the LGS source is sampled in 1D. For a given sub-aperture, each of the sampling points is re-imaged on the Ingot main plane, including the effect of the local tilt, induced by the atmosphere in the portion of the wavefront selected by that sub-aperture. The re-imaged point is then convolved with the disk image, corresponding to the given sub-aperture, and summed to all the images, produced by the points sampling the LGS axis. Differential intensities can be applied to the disks, so that the Sodium layer density profile is reproduced. Figure 10 shows the resulting LGS image, produced on the Ingot main plane, for four different sub-apertures. Each of the LGS images is finally filtered by the corresponding Ingot projection mask, to compute the fraction of light illuminating each of the three faces of the Ingot. The retrieved relative illumination is the applied to the corresponding three sub-apertures in the three simulated pupil images.

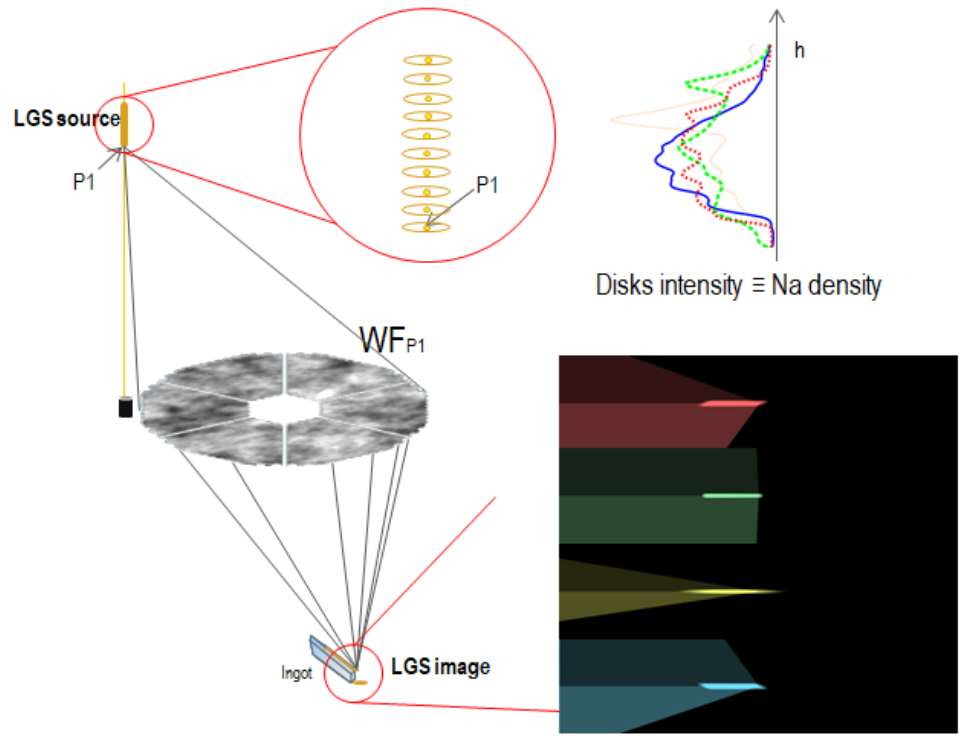

Figure 10. Example of LGS images and Ingot projections on the Ingot main plane for four different sub-apertures. Atmospheric aberration and differential Sodium layer density profile not applied in this example.

This approach allows to include in the simulation, without further assumptions:

- Sodium layer density profile,
- atmospheric aberration, which is different for each point sampling the LGS, and
- Gaussian profile of the LGS in the horizontal plane.

## 5. CONCLUSION

We reported the current scheme adopted for I-WFS performance simulation, which takes advantage of a hybrid approach, to minimize the number of approximations to be done (e.g. prism vertex angle is not close to zero) and includes as realistic as possible information, anyway keeping the computing power and time needed to perform the simulation under control.

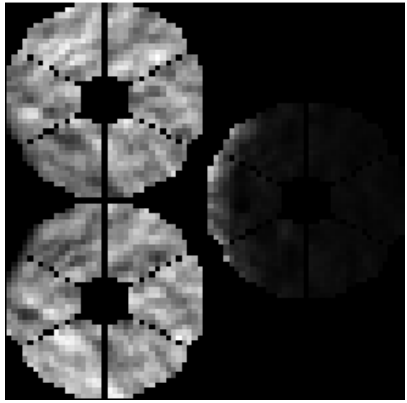

Figure 11. Ingot WFS pupils with atmospheric aberration (Von Karman) included.

Figure 11 shows an example of resulting pupil images, obtained with the presented method. The pupils present the uneven illumination, which is an Ingot WFS characteristic that can be easily recognized also when no incoming aberration is present.

This illumination pattern is due to the Ingot prism edges projections, as seen from different sub-apertures, which are not parallel to X-Y axis (as shown in Figure 9). This translates into an excess/lack of light in corresponding parts of the pupils, reported in Figure 11.

The presented approach is demonstrating to satisfy stability and repeatability needs of I-WFS simulations, and properly working within the wavefront reconstruction process, presented in Portaluri et al. [4], and tested with first laboratory experiments, reported in Di Filippo et al. [10].